\documentclass[aps,preprint,epsfig,rotate]{revtex4}
\usepackage{graphicx}
\usepackage{bm}
\usepackage{epsfig}
\input epsf
\begin{document}
%\begin{doublespace}

\title{Positron annihilation in the MuPs system}
 \author{Alexei M. Frolov}
 \email[E--mail address: ]{afrolov@uwo.ca}

\affiliation{Department of Chemistry\\
 University of Western Ontario, London, Ontario N6H 5B7, Canada}

\date{\today}

\begin{abstract}

The life-time of the four-body atomic system MuPs ($\mu^{+} e^{-}_2 e^{+}$
or muonium-positronium) against positron annihilation has been evaluated as
$\tau = \frac{1}{\Gamma} \approx 4.076479 \cdot 10^{-10}$ $sec$. Various
annihilation rates for MuPs are determined to a good numerical accuracy,
e.g., $\Gamma_{2 \gamma} \approx$ 2.446469$\cdot 10^{9}$ $sec^{-1}$,
$\Gamma_{3 \gamma} \approx$ 6.62793$\cdot 10^{6}$ $sec^{-1}$, $\Gamma_{4
\gamma} \approx$ 3.61677$\cdot 10^{3}$ $sec^{-1}$, $\Gamma_{5 \gamma}
\approx$ 6.32969 $sec^{-1}$. The hyperfine structure splitting for the 
ground state in the MuPs system has also been evaluated as $\Delta$ = 
23.078 $MHz$.

\end{abstract}
\maketitle
\newpage

In our earlier study \cite{FrWa2010} we have considered the bound states in
the positronium hydrides ${}^{\infty}$HPs, TPs, DPs, ${}^{1}$HPs and MuPs
($\mu^{+} e^{-}_2 e^{+}$). In this work we report the improved results
obtained recently for four-body muonium-positronium system MuPs. In these
calculations we could not improve the total (and binding) energies of the
MuPs system, but some expectation values of interparticle delta-functions
have been changed noticeably (in contrast with the positronium hydrides!).
These changes in delta-functions lead to variations in the computed rates of
positron annihilation. The improved annihilation rates can directly be
compared with the experimental values, e.g., with the life-time of MuPs
against positron annihilation.

In general, for the four-body MuPs system ($\mu^{+} e^{-}_2 e^{+}$) one can
find many interesting physical problems to investigate (see, e.g.,
\cite{FrWa2010} and references therein). The MuPs system has only one bound
${}^1S_e$-state and this state is the electron singlet state, i.e. the two 
electron spins are oriented in the opposite directions (otherwise, the MuPs
system is not bound). In this study we restrict ourselves to the discussions 
of the positron annihilation and hyperfine structure splitting. The 
muonium-positronium conversion and other interesting problems \cite{FrWa2010} 
are not considered here.

The Hamiltonian of the four-body $\mu^{+} e^{-}_2 e^{+}$ system is written
in the form (in atomic units $\hbar = 1, m_e = 1, e = 1$):
\begin{eqnarray}
 H = -\frac{1}{2 m_{\mu}} \Delta_{1} -\frac{1}{2} \Delta_{2}
     -\frac{1}{2} \Delta_{3} -\frac{1}{2} \Delta_{4} + \frac{1}{r_{12}}
     - \frac{1}{r_{13}} - \frac{1}{r_{14}} - \frac{1}{r_{23}}
     - \frac{1}{r_{24}} + \frac{1}{r_{34}} \label{eq1}
\end{eqnarray}
where the notation 1 designates the positively charged muon $\mu^{+}$,
the notation 2 (or +) means the positron, while 3 (or -) and 4 (or -) stand
for electrons. This system of notations will be used everywhere below in our
study. By solving the corresponding Schr\"{o}dinger equation $H \Psi = E
\Psi$ one can determine the wave function of all bound states in the MuPs
system. In reality, there is only one bound (ground) state in the
muonium-positronium MuPs. The total energy of this ground state is $\approx$
-0.78631730(15) $a.u.$ \cite{FrWa2010}. The known accurate wave functions
$\Psi$ allows one to obtain the expectation values of many operators,
including the electron-positron and muon-positron delta-functions,
respectively. In turn, such delta-functions are needed to evaluate the
positron annihilation rate(s) in the MuPs system and predict the hyperfine
structure splitting. Note that all numerical calculations in this paper
have been performed with the use of variational expansion in four-dimensional
gaussoids in relative (interparticle) coordinates $r_{ij}$ \cite{KT}. This
expansion was proposed more than 30 years ago in \cite{KT} (see also earlier
references therein). The explicit form of this expansion for the ground 
$S(L = 0)-$state in the four-body $MuPs$ system is
\begin{eqnarray}
 \Psi_{L=0} = (1 + {\cal P}_{34}) \sum_{k=1}^N
 C_k \cdot exp( -\alpha^{(k)}_{12} r^{2}_{12} -\alpha^{(k)}_{13} r^{2}_{13}
 -\alpha^{(k)}_{23} r^{2}_{23} -\alpha^{(k)}_{14} r^{2}_{14}
 -\alpha^{(k)}_{24} r^{2}_{24} -\alpha^{(k)}_{34} r^{2}_{34}) \label{Gaus}
\end{eqnarray}
where $C_k$ are the linear coefficients (or linear variational parameters),
while $\alpha^{(k)}_{ij}$ are the optimized non-linear parameters. The
notation ${\cal P}_{34}$ means the permutation operator for the two identical
particles 3 and 4 (electrons) in this system. In our calculations we have 
constructed the trial wave functions with $N$ = 600, 800, 1000 and 1400 in  
Eq.(\ref{Gaus}). The use of these wave functions with the carefully optimized
non-linear parameters allows one to obtain the accurate expectation values
all operators needed in this study. More details about the variational 
expansion, Eq.(\ref{Gaus}), and optimization of the non-linear parameters in 
it can be found in our earlier papers (see, e.g., \cite{FrWa2009} and 
references therein).  

Annihilation of electron-positron pairs in the MuPs system is a very
interesting process which can be observed experimentally. The largest
annihilation rates correspond to the two- and three-photon annihilation. For
the two-photon annihilation in \cite{FrWa2010} we have obtained the
following formula
\begin{eqnarray}
 \Gamma_{2 \gamma}({\rm MuPs}) = 2 \pi \alpha^4 c a^{-1}_0 \Bigl[ 1 -
 \frac{\alpha}{\pi} \Bigl( 5 - \frac{\pi^2}{4} \Bigr)\Bigr] \langle
 \delta({\bf r}_{+-}) \rangle = 100.3456053781 \cdot 10^{9} \langle
 \delta_{+-} \rangle \; sec^{-1} \; \; \; . \label{An2g}
\end{eqnarray}
where $\langle \delta_{+-} \rangle$ is the expectation value of the
electron-positron delta-function determined for the ground state in the
MuPs system. The formula for the three-photon annihilation rate $\Gamma_{3
\gamma}({\rm MuPs})$ takes the form
\begin{eqnarray}
 \Gamma_{3 \gamma}({\rm MuPs}) = 2 \frac{4 (\pi^2 - 9)}{3} \alpha^5 c
 a^{-1}_0 \langle \delta({\bf r}_{+-}) \rangle = 2.718545954 \cdot
 10^8 \langle \delta_{+-} \rangle \; sec^{-1}
\end{eqnarray}
In these formulas and everywhere below $\alpha$ is the fine structure
constant, $c$ is the speed of light in vacuum and $a_0$ is the Bohr radius.
Below, the numerical values of these constants have been taken from
\cite{NIST} and \cite{CRC}.

The rates of the four- and five-photon annihilations of the
electron-positron pairs in the MuPs system are related with the $\Gamma_{2
\gamma}({\rm MuPs})$ and $\Gamma_{3 \gamma}({\rm MuPs})$, respectively, by
the following approximate relations \cite{PRA83}
\begin{equation}
 \Gamma_{4 \gamma}({\rm MuPs}) \approx 0.274
 \Bigl(\frac{\alpha}{\pi}\Bigr)^2 \Gamma_{2 \gamma}({\rm MuPs})
 \approx 1.478364 \cdot 10^{-6} \cdot \Gamma_{2 \gamma}({\rm MuPs})
 \label{e4a}
\end{equation}
and
\begin{equation}
 \Gamma_{5 \gamma}({\rm MuPs}) \approx 0.177
 \Bigl(\frac{\alpha}{\pi}\Bigr)^2 \Gamma_{3 \gamma}({\rm MuPs})
 \approx 9.550018 \cdot 10^{-7} \cdot \Gamma_{3 \gamma}({\rm MuPs})
 \label{e5a}
\end{equation}

To a very good accuracy one can evaluate the total annihilation rate of the
MuPs system by the following sum $\Gamma \approx \Gamma_{2 \gamma} +
\Gamma_{3 \gamma} + \Gamma_{4 \gamma} + \Gamma_{5 \gamma} \approx \Gamma_{2
\gamma} + \Gamma_{3 \gamma} \approx 1006.174599735 \cdot 10^{8} \langle
\delta_{+-} \rangle$ $sec^{-1}$ $\approx 2.4530973 \cdot 10^{9}$
$sec^{-1}$. In other words, the knowledge of accurate values of the
$\Gamma_{2 \gamma}$ and $\Gamma_{3 \gamma}$ annihilation rates is
sufficient to predict the total life-time of the MuPs system against
positron annihilation $\tau = \frac{1}{\Gamma} \approx 4.076479 \cdot
10^{-10}$ $sec$.

In addition to the two-, three-, four- and five-photon annihilations of the
electron-positron pair the one-photon and zero-photon annihilations may also
play an important role in some applications. In general, these rates are
very small. Note that these rates have been discussed (and evaluated) in our
earlier studies (see, e.g., \cite{FrWa2010} and references therein). In
particular, the closed formula for the zero-photon annihilation rate
$\Gamma_{0 \gamma}$ takes the form
\begin{eqnarray}
 \Gamma_{0 \gamma} = \xi \frac{147 \sqrt{3} \pi^3}{2} \cdot \alpha^{12}
 (c a_0^{-1}) \cdot \langle \delta_{\mu^{+}+--} \rangle =  5.0991890 \cdot
 10^{-4} \cdot \xi \cdot \langle \delta_{\mu^{+}+--} \rangle \; \; \;
 sec^{-1} \label{0phot}
\end{eqnarray}
where $\langle \delta_{\mu^{+}+--} \rangle$ is the expectation value of the
four-particle delta-function in the ground state of muonium-positronium
(MuPs). The numerical value of $\langle \delta_{\mu^{+}+--} \rangle$ is the
probability to find all four particles at one volume with the radius $\alpha
a_0$. The unknown (dimensionless) factor $\xi$ has the numerical value close
to unity. The expectation value of the four-particle delta-function
determined in our calculations is $\approx 1.76789 \cdot 10^{-4}$ (in
$a.u.$). From here one finds that $\Gamma_{0 \gamma}$(MuPs) $\approx
9.0320(10) \cdot 10^{-8} \xi$ $sec^{-1}$. For approximate evaluations we
can assume that the factor $\xi$ equals unity. In this case one finds that
$\Gamma_{0 \gamma}$(MuPs) $\approx 9.0320(10) \cdot 10^{-8}$ $sec^{-1}$.

The one-photon annihilation of the electron-positron pair in MuPs can be
considered as a regular two-photon annihilation, but one of the two emitted
photons is absorbed either by the remaining electron $e^{-}$, or by the muon
$\mu^{+}$. In the
first case one can observe the emission of the fast electron. The
probability of this process is given by the formula
\begin{eqnarray}
 \Gamma^{(1)}_{1 \gamma} = \frac{64 \pi^2}{27} \cdot \alpha^{8}
 (c a_0^{-1}) \cdot \langle \delta_{+--} \rangle = 1.066420947 \cdot 10^3
 \cdot \langle \delta_{+--} \rangle \; \; \; sec^{-1} ,
\end{eqnarray}
where $\langle \delta_{+--} \rangle = \langle \delta({\bf r}_{+-})
\delta({\bf r}_{--}) \rangle)$ is the expectation value of the triple
electron-positron delta-function determined for the ground state of the MuPs
system. Its numerical value is the probability to find all three
corresponding particles at one spatial point with spatial radius $\alpha a_0
\approx \frac{a_0}{137}$. Our best numerical treatment to-date for the
$\langle \delta_{+--} \rangle$ value gives $\approx 3.68652 \cdot 10^{-4}$,
and therefore, $\Gamma^{(1)}_{1 \gamma} \approx$ 3.9314(20)$\cdot 10^{-1}$
$sec^{-1}$ for the ground state in the MuPs system.

Analysis of the second one-photon annihilation in MuPs is significantly more
complicated (see discussion in \cite{FrWa2010}). To evaluate the
corresponding annihilation rate $\Gamma^{(2)}_{1 \gamma}$ one needs
undertake an extensive QED consideration \cite{AB}. Such an analysis is also
required to determine the accurate value of the factor $\xi$ in
Eq.(\ref{0phot}). Here we shall not discuss these problems.

Now, let us discuss the hyperfine structure splitting in the MuPs system.
The hyperfine structure splitting in the MuPs system is written in the form
\cite{FrWa2010}
\begin{equation}
 a = 14229.1255 \cdot \langle \delta_{\mu^{+} e^{+}} \rangle \label{spl31}
\end{equation}
To obtain this formula we have used the following values for the muon mass
$m_{\mu}$ and for the factors $g_{+}$ and $g_{\mu}$ \cite{NIST},
\cite{CRC}:
\begin{eqnarray}
 m_{\mu} = 206.768264 m_e \; \; \; , \; \; \; g_{+} = -2.0023193043718 \;
 \; \; \; \; \; g_{\mu} = -2.0023318396
\end{eqnarray}
where $m_e$ is the electron/positron mass at rest. By using our improved
expectation value for the muon-positron delta-function $\langle
\delta_{\mu^{+} e^{+}} \rangle \approx 1.6218815 \cdot 10^{-3}$ $a.u.$, one
finds that the constant in Eq.(\ref{spl31}) is $a \approx 23.078$ $MHz$.
This is the energy difference between the state with $J = 0$ and $J = 1$,
where the notation $J$ stands the total spin of the muon-positron pair.

Thus, we have considered the positron annihilation in the MuPs system
($\mu^{+} e^{-}_2 e^{+}$ or muonium-positronium). The two-, three-, four-
and five-photon annihilation rates of the MuPs system are determined to high
numerical accuracy (see Table I). The rate of zero-photon annihilation
$\Gamma_{0 \gamma}$(MuPs) and first one-photon annihilation rate
$\Gamma^{(1)}_{1 \gamma}$(MuPs) have also been estimated (see Table I). To
evaluate zero-photon annihilation rate $\Gamma_{0 \gamma}$(MuPs) to better
accuracy and obtain the explicit formula for the second one-photon
annihilation rate $\Gamma^{(1)}_{1 \gamma}$(MuPs) one needs to perform an
extensive QED analysis. We also determine the hyperfine structure splitting
between singlet $J = 0$ and triplet $J = 1$ spin states of the muon-positron
pair in MuPs. By using our expectation value for the $\delta_{\mu^{+}+}$
delta-function (see Table I) we have found that the hyperfine splitting
in the ground state of the MuPs system is $\approx$ 23.078(10) $MHz$.

Note that the annihilation of electron-positron pairs in various light
atomic systems was considered in many earlier works (see, e.g., 
\cite{Ho}, \cite{Fro1} and references therein). The positron annihilation
in the MuPs system has been discussed earlier in \cite{FrWa2010}. The 
expectation values for the delta-functions obtained in this paper for the 
MuPs system have better overall accuracy than analogous values evaluated in 
\cite{FrWa2010}. Numerical evaluations of some relativistic corrections of 
the lowest order in positronium hydrides and MuPs can be found in 
\cite{Ho99}.  

\newpage
  \begin{table}[tbp]
   \caption{The expectation values of some delta-functions in the four-body 
            MuPs system and annihilation rates $\Gamma_{n \gamma}$ (in 
            $sec^{-1}$) of the $n$-photon annihilation ($n$ = 2, 3, 4, 5, 1, 
            0) of the electron-positron pair.}
     \begin{center}
     \begin{tabular}{lllll}
       \hline \hline
 $\langle \delta_{+-} \rangle$ &
 $\Gamma_{2 \gamma}$ & $\Gamma_{3 \gamma}$ &
 $\Gamma_{4 \gamma}$ & $\Gamma_{5 \gamma}$ \\
    \hline
 2.4380433$\cdot 10^{-2}$ & 2.446469$\cdot 10^{9}$ &
 6.62793$\cdot 10^{6}$ & 3.61677$\cdot 10^{3}$ &
 6.32969$\cdot 10^{0}$ \\
     \hline
     \hline
 $\langle \delta_{+--} \rangle$ &
 $\Gamma^{(1)}_{1 \gamma}$ &
 $\langle \delta_{\mu^{+}+--} \rangle$ &
 $\Gamma_{0 \gamma}$ & $\langle \delta_{\mu^{+}+} \rangle$ \\
     \hline
 3.68525$\cdot 10^{-4}$ &
 3.9314$\cdot 10^{-1}$ &
 1.76789$\cdot 10^{-4}$ &
 9.0148$\cdot 10^{-8}$ &
 1.621887$\cdot 10^{-3}$ \\
    \hline \hline
  \end{tabular}
  \end{center}
  \end{table}
\end{document}